\documentclass[12pt]{revtex4}

\usepackage{epsfig}
\usepackage{graphicx}
\usepackage{color}
\usepackage{multirow}
\usepackage{amssymb}
\usepackage{color}

\begin{document}

\title{Calculated low-energy electron-impact vibrational excitation cross sections for CO$_2$ molecule}

\author{
V. Laporta$^{1,2,*}$\footnote[0]{$^*$ vincenzo.laporta@nanotec.cnr.it},
J. Tennyson$^{2}$ and
R. Celiberto$^{3,1}$
}
\affiliation{$^1$Istituto di Nanotecnologia, CNR, 70126 Bari, Italy}
\affiliation{$^2$Department of Physics and Astronomy, University College London, London WC1E 6BT, UK}
\affiliation{$^3$Dipartimento di Ingegneria Civile, Ambientale, del Territorio, Edile e di Chimica, Politecnico di Bari, 70126 Bari, Italy}

\begin{abstract}
Vibrational-excitation cross sections of ground electronic state of carbon dioxide molecule by electron-impact through the CO$_2^-(^2\Pi_u)$ shape resonance is considered in the separation of the normal modes approximation. Resonance curves and widths are computed for each vibrational mode.  The calculations assume decoupling between normal modes and employ the local complex potential model for the treatment of the nuclear dynamics, usually adopted for the electron-scattering involving diatomic molecules. Results are presented for excitation up to 10 vibrational levels in each mode and comparison with data present in the literature is discussed.
\end{abstract}

\maketitle

One of the technological problems, connected with strategies for
reduction of the global warming coming from the greenhouse effect
produced by carbon dioxide, is represented by the capture at source
and storage of the CO$_2$ gas, mainly based on the plasmolysis process
leading to the splitting of CO$_2$  into CO molecules and atomic or molecular
oxygen~\cite{Taylan_Berberoglu_2015, Goede_et_al_2014}. The efficiency
of the dissociation processes is strongly determined by the
vibrational activation of the molecule. Models of CO$_2$
plasmas, aimed to optimize and clarify this chemical conversion, have
recently been constructed~\cite{doi:10.1021/acs.jpca.6b01154,
  0963-0252-24-1-015024, doi:10.1021/jp408522m, 0963-0252-23-4-045004,
  doi:10.1021/jp509843e}. The main limitation of these models is
the lack of information on electron-impact cross
sections or rate coefficients for collisions inducing vibrational
transitions in CO$_2$ molecules; as result modellers usually resort to
estimated rates or approximate scaling-laws~\cite{0963-0252-23-4-045004}.

In order to fill this void, in this Letter we present a preliminary
data set of electron-impact cross sections for vibrational
excitation of ground electronic state of carbon dioxide molecule useful in plasma kinetic
modeling. The cross sections show two distinctive features observed
experimentally: a $^2\Pi_u$ shape resonance around 3.8
eV~\cite{PhysRevA.15.2186, 0022-3700-10-18-034, PhysRevLett.87.033201}
and, at energies below 2 eV, an enhancement due to the presence of the
$^2\Sigma_g^+$ symmetry virtual state~\cite{PhysRevLett.80.1873,jt238,
  PhysRevA.64.040701}. Both phenomena are explained in terms of a
temporary CO$_2^-$ system. For a general review on this topics see Itikawa paper~\cite{itikawa:749} and references therein.

We present here the cross sections for the following process:
\begin{equation}
e + \mathrm{CO}_2(\textrm{X}\,^1\Sigma_g^+; v)\rightarrow \mathrm{CO}_2^-(^2\Pi_u) \rightarrow e + \mathrm{CO}_2(\textrm{X}\,^1\Sigma_g^+; v') \label{process}\,,	
\end{equation}
which occurs through the formation of the shape resonance generated by
the electronic state $^2\Pi_u$ of the $\mathrm{CO}_2^-$ ion. CO$_2$ in
its ground electronic state, $\textrm{X}\,^1\Sigma_g^+$, is a linear molecule
with C--O equilibrium distance $R_{eq}=2.19$~$a_0$
characterized by three normal modes of vibration, denoted in the
following by $v=(\nu_1,\nu_2,\nu_3)$ and referred to, respectively, as
the symmetric stretching, bending mode (doubly degenerate) and
asymmetric stretching.

\textcolor{green}{In principle, scattering involving polyatomic molecules needs a
multidimensional treatment of the potential energy surface and of the
nuclear motion, in order to take into account the non-adiabatic
coupling between different vibrational modes \cite{estrada:152}. However, as we are
limiting ourself to the lowest vibrational levels, where the potential
energy is approximatively harmonic, is possible to adopt the
assumption of separation of the modes and split the CO$_2$ potential
into three one-dimensional independent modes.} This allows one to
compute the cross sections employing the local model of resonant
collisions as formulated for diatomic vibrational
excitation~\cite{0963-0252-21-5-055018,0963-0252-23-6-065002, PhysRevA.91.012701}. \textcolor{green}{In the uncoupled vibrational mode approximation, each mode is considered as independent. This implies that the scattering processes involves one mode only and does not affect at all the other two.} Preliminary results for the
symmetric stretch mode only, were given
previously~\cite{Celiberto_et_al_OPPJ_2014}. Here we present
calculations on all the three normal modes of the molecule, for $0\leq
\nu_i\leq \nu_i'\leq 10$ ($i=1,2,3$), and for electron collision
energies from the threshold up to 10~eV.

A peculiar aspect of the the doubly degenerate $^2\Pi_u$ symmetry of
CO$_2^-$ ion is that it splits, upon bending, into two (Renner-Teller)
$^2A_1$ and $^2B_1$ components, no longer degenerate, due to the
breaking of linear geometry ($D_{2h}$ symmetry to $C_{2v}$ symmetry of
bending mode) \cite{PhysRevA.65.032716, PhysRevA.67.042708,
  Herzberg_III}. A second aspect, that derives from the stretch-bend
coupling possible in polyatomic molecules, is an accidental
degeneracy of vibrational levels belonging to different modes, known
as Fermi resonance \cite{Herzberg_III,itikawa:749}. In the case of
CO$_2$ a quasi-degeneracy occurs between the pure stretch $(100)$ and
the pure bending $(020)$ levels (Fermi dyads) that result in a near 50:50 mixing of
the two states which is well-known experimentally
\cite{PhysRevLett.87.033201}. Here we neglect this stretch-bend
coupling; it is at least arguable that including it will only result
in redistribution of flux in the excitation cross sections rather
than radically different excitation rates.

\begin{figure}
\begin{tabular}{ccc}
Symmetric stretching & Bending mode & Asymmetric stretching\\
\includegraphics[scale=.55]{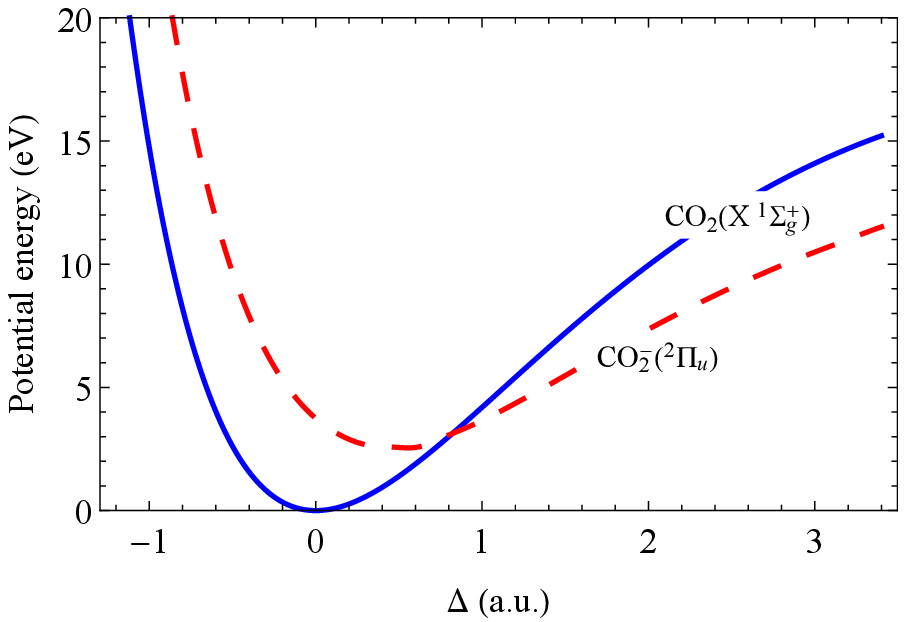}   & \includegraphics[scale=.55]{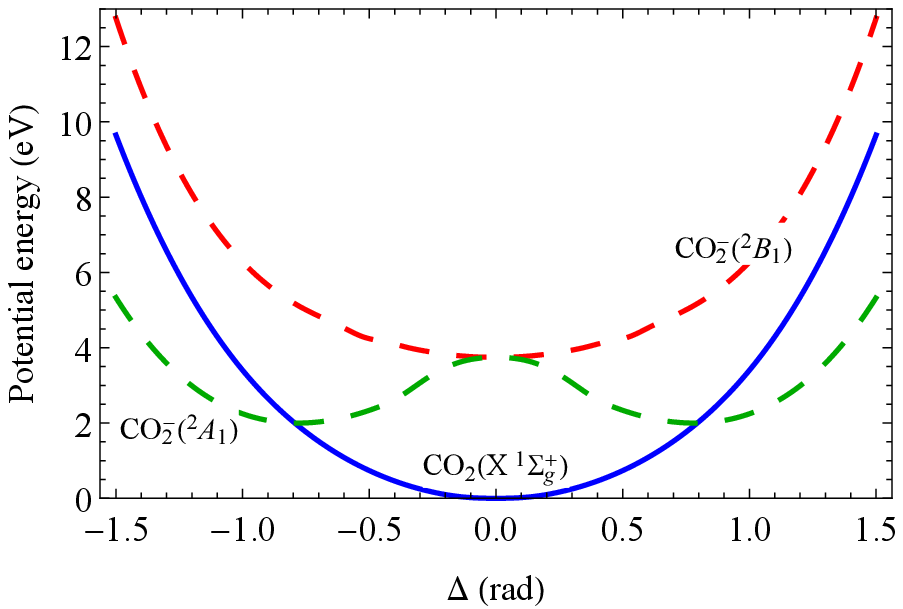}   & \includegraphics[scale=.55]{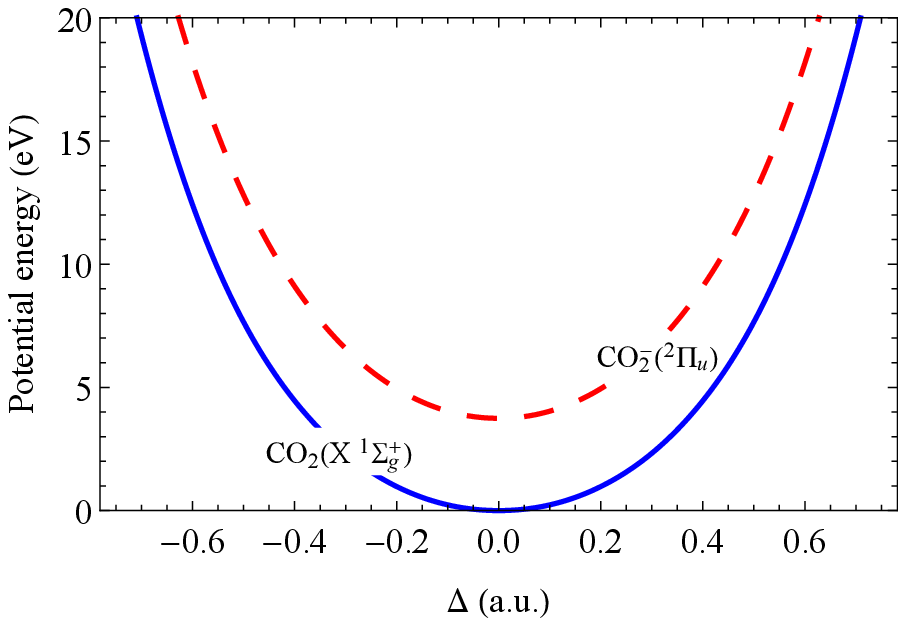}\\
\includegraphics[scale=.55]{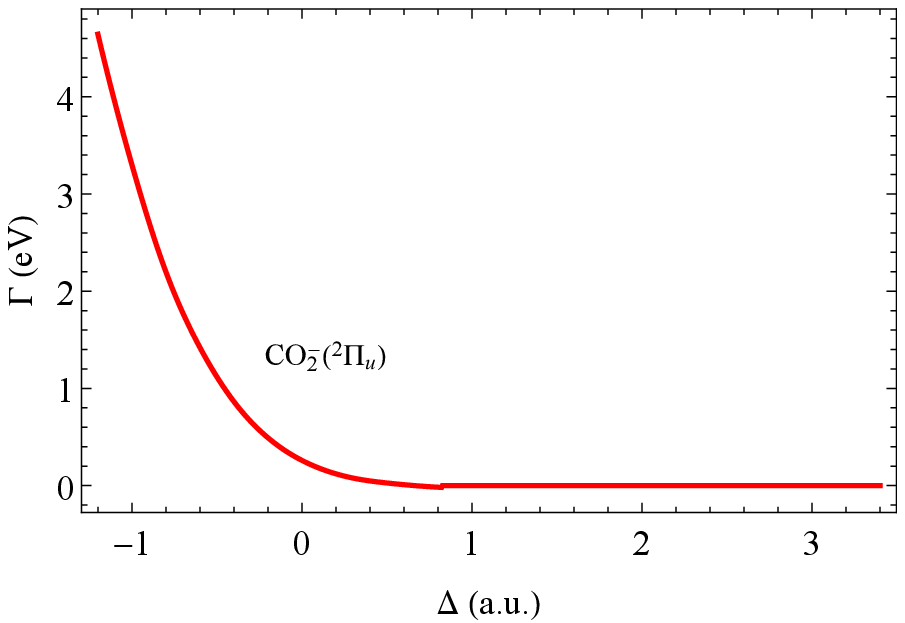} & \includegraphics[scale=.55]{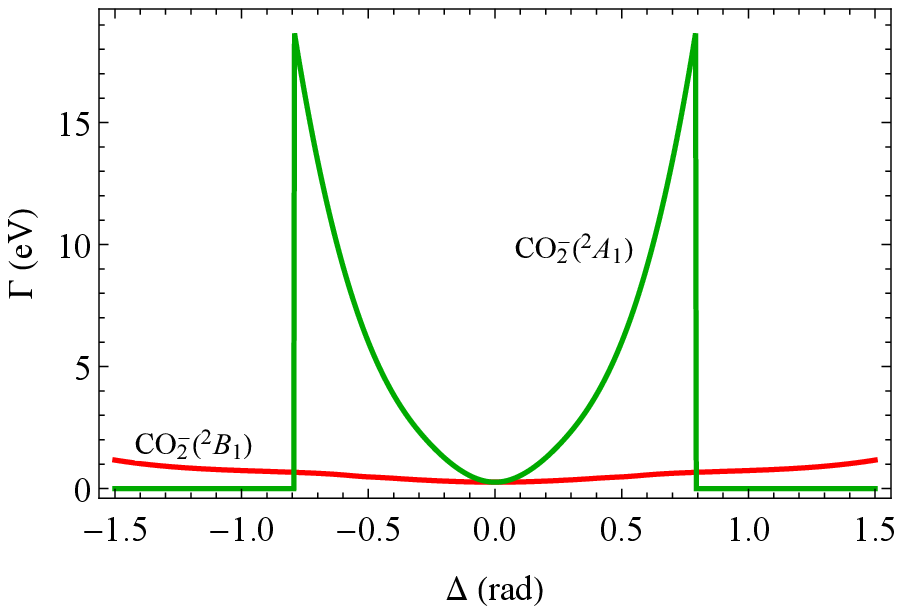} & \includegraphics[scale=.55]{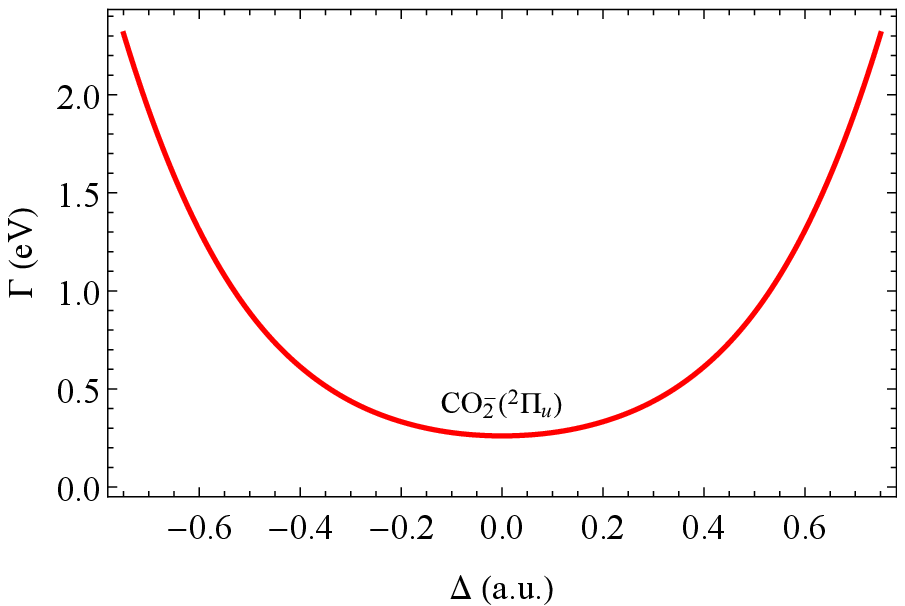}
\end{tabular}
\caption{Upper panels: Potential energy curves for the electronic ground electronic state $\textrm{X}\,^1\Sigma^+_g$ of CO$_2$ (full blue line) and for the resonant state CO$_2^-$ (dashed red and green lines) for the three normal mode symmetric, bending and asymmetric. Lower panels: the corresponding resonance widths. $\Delta$ represents the displacement from the equilibrium geometry. \label{fig:CO2pes}}
\end{figure}
The CO$_2$ potential energy curves were computed using the \emph{ab-initio} quantum chemistry code MOLPRO
\cite{MOLPRO_brief}, by adopting a aug-cc-pVQZ basis set and the
coupled-cluster (CCSD(T)) and MCSCF models. Scattering calculations were performed
using the UK polyatomic R-matrix codes \cite{Tennyson_PR_2010,jt518}. A static exchange plus polarization (SEP) model,
and the same basis used for CO$_2$, were utilised to calculate the
complex potential energy curve for CO$_2^-$. The \textit{R}-matrix
calculations were performed on a grid of fixed internuclear distances.
The position and width of the resonant state were then calculated by
fitting the corresponding eigenphases sum with a Breit-Wigner function
\cite{Tennyson1984421}.

The potential energy curves for CO$_2$ and CO$_2^-$ species and for
the three normal modes, are shown in Fig.~\ref{fig:CO2pes} along with
the corresponding resonance widths $\Gamma$. These curves are plotted
as a function of the atomic displacements, $\Delta$, calculated with
respect to the equilibrium geometry for the three modes. Table
\ref{table:CO2viblev} reports the vibrational energy levels
$\epsilon_{\nu_i}$ for each normal mode ($i=1,2,3$).

\begin{table}
\begin{tabular}{cccc}
\hline
        & Symmetric stretching& Bending mode & Asymmetric stretching\\
$\nu_i$ & $i=1$ & $i=2$ & $i=3$\\
\hline
0  &  0.     & 0.     & 0.     \\
1  &  0.1676 (0.172) & 0.0764 (0.082)& 0.2973 (0.291) \\
2  &  0.3345 & 0.1585 (0.159) & 0.5996 \\
3  &  0.5007 & 0.2415 & 0.9052 \\
4  &  0.6661 & 0.3246 & 1.2138 \\
5  &  0.8309 & 0.4081 & 1.5252 \\
6  &  0.9949 & 0.4919 & 1.8396 \\
7  &  1.1583 & 0.5760 & 2.1569 \\
8  &  1.3208 & 0.6604 & 2.4771 \\
9  &  1.4827 & 0.7450 & 2.8000 \\
10 &  1.6439 & 0.8299 & 3.1257 \\
\hline\hline
$\epsilon_{\nu_i=0}$ & 0.0840 & 0.0357 & 0.1469 \\
\hline
\end{tabular}
\caption{CO$_2$ vibrational levels $\epsilon_{\nu_i}$ in the three normal modes referred to the energy of the corresponding ground vibrational levels whose value $\epsilon_{\nu_i=0}$, with respect to the minimum of the potential energy curve, is given in the last row. In parenthesis are shown the experimental values from Ref. \cite{Herzberg_III}. All entries are in eV. \label{table:CO2viblev}}
\end{table}

The resonant cross section for the process in (\ref{process}) of a
single normal mode $\nu$ has been calculated \cite{PhysRevA.67.042708,
  0963-0252-21-5-055018,Celiberto_et_al_OPPJ_2014}, as a function of
the electron energy $\epsilon$, by:
\begin{equation}
\sigma_{\nu\to\nu'}(\epsilon)=\frac{4\pi^3}{k^2}|T_{\nu\nu'}|^2 \label{xsec}\,,
\end{equation}
where $k$ is the incoming electron momentum and the scattering matrix, $T_{\nu\nu'}$, as a function of the total energy $E=\epsilon+\epsilon_\nu$, is defined as:
\begin{equation}
T_{\nu\nu'}(E) = \langle \chi_{\nu'}^*|  \mathcal{V} (\mathcal{H} - E)^{-1}  \mathcal{V} | \chi_\nu \rangle \label{t-matrix}\,.
\end{equation}
In Eq.(\ref{t-matrix}), $\chi_{\nu(\nu')}$ is the wave function of the
initial (final) vibrational level; $\mathcal H = T_N + V^- -\frac i
2\Gamma$ is the Hamiltonian of the system, where $T_N$ is the kinetics
energy operator and $V^- -\frac i 2\Gamma$ is the \textit{ complex optical
 potential} of the resonance; $\mathcal{V}=\sqrt{\Gamma/2}$
is the bound-continuum coupling matrix. Full details about the theoretical model can be found in the paper \cite{0034-4885-31-2-302} and references therein. The cross section involving different modes $(\nu_1,\nu_2,\nu_3)$ is given by the coherent sum over the cross sections for the three normal modes $\nu_i$.

\begin{figure}
\begin{tabular}{ccc}
Symmetric stretching & Bending mode & Asymmetric stretching\\
\includegraphics[scale=.6]{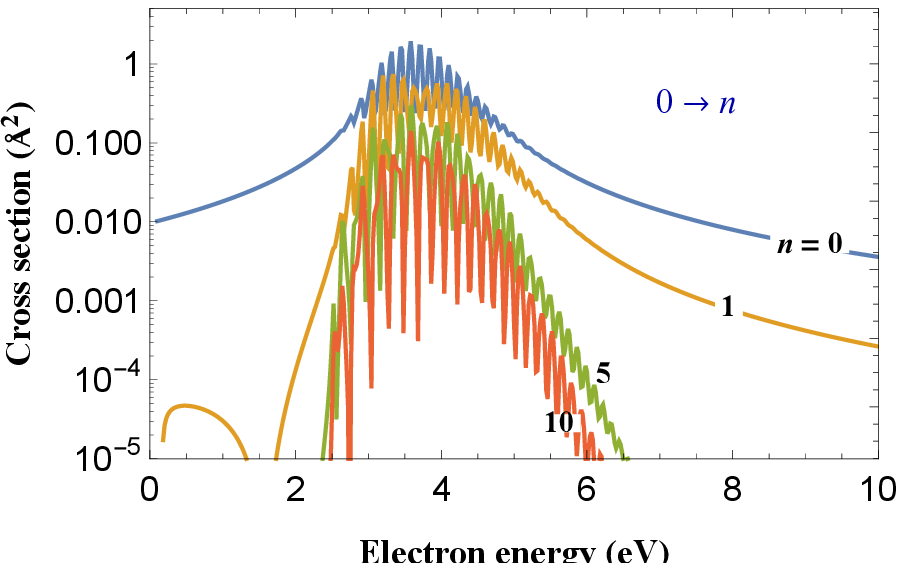}&\includegraphics[scale=.6]{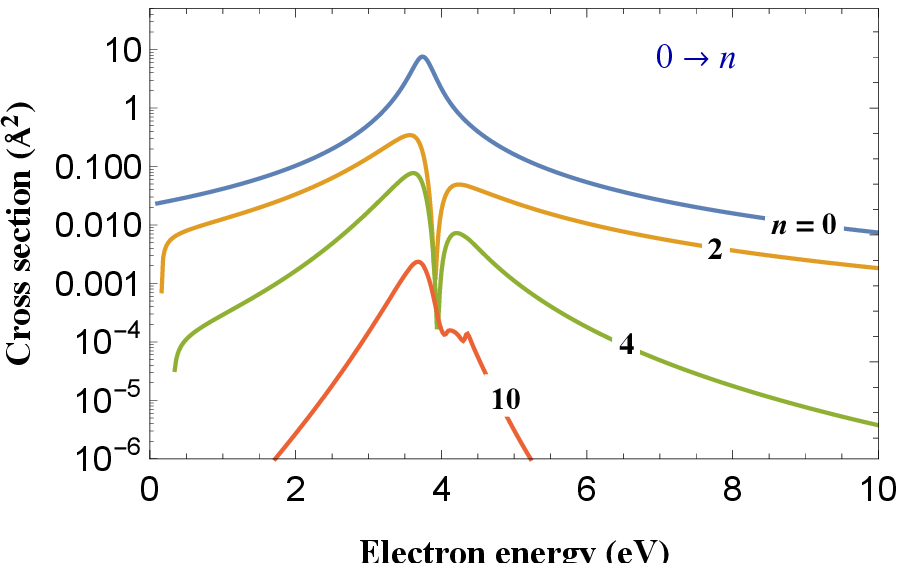}&\includegraphics[scale=.6]{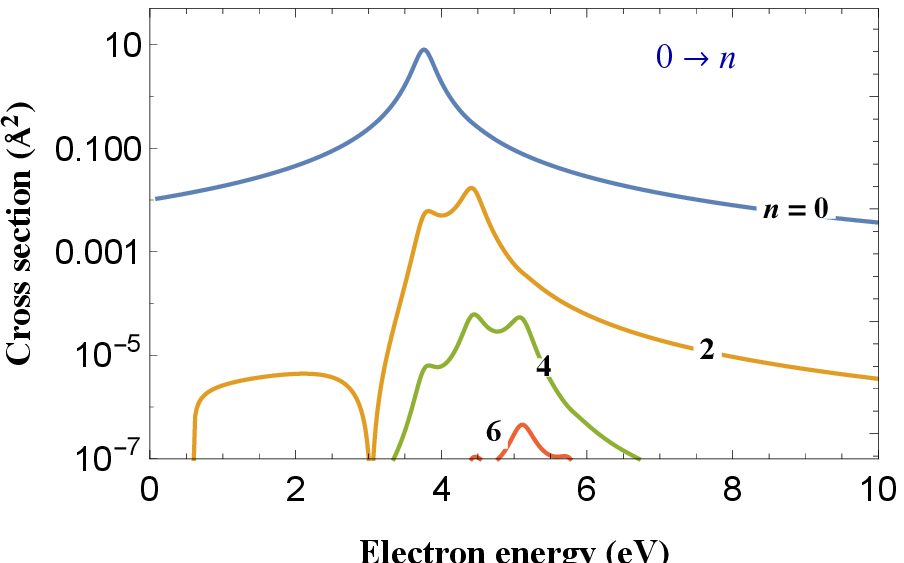}\\
\includegraphics[scale=.6]{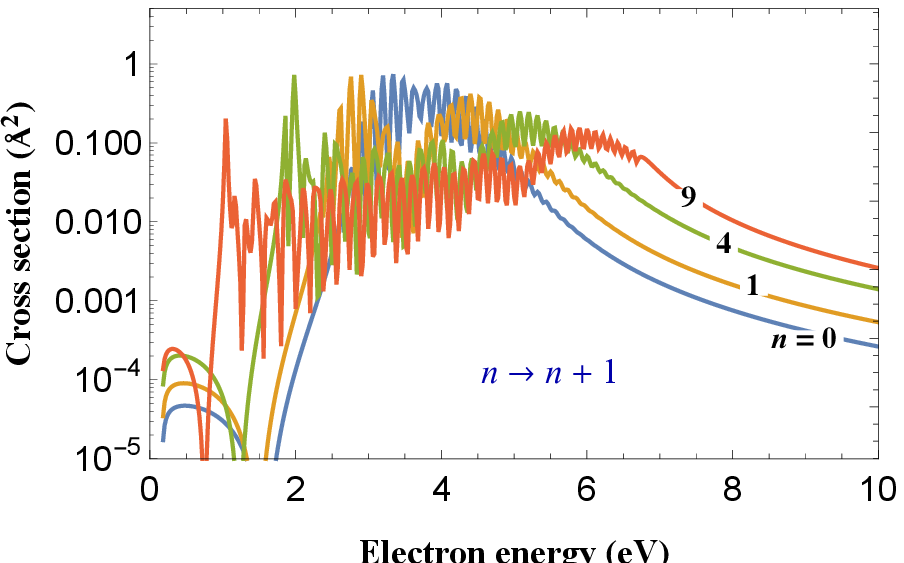}&\includegraphics[scale=.6]{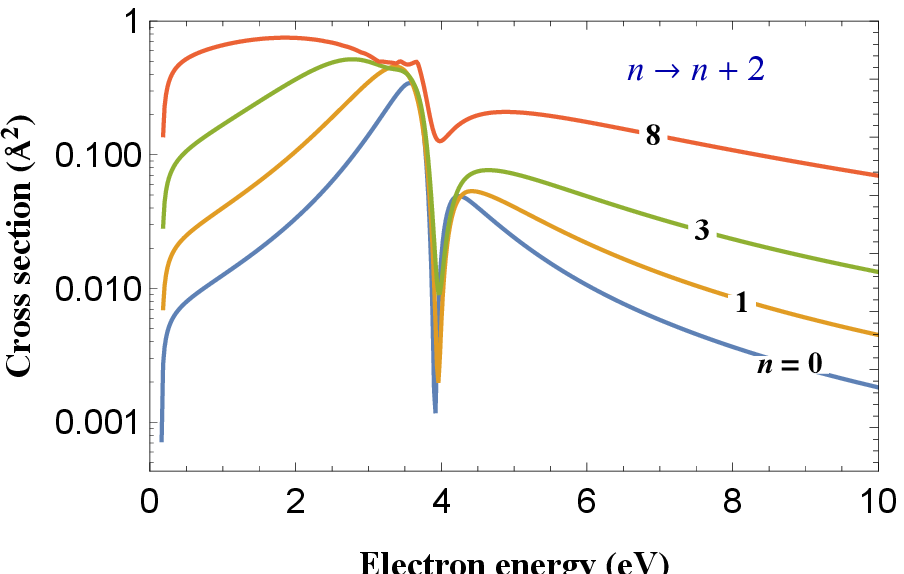}&\includegraphics[scale=.6]{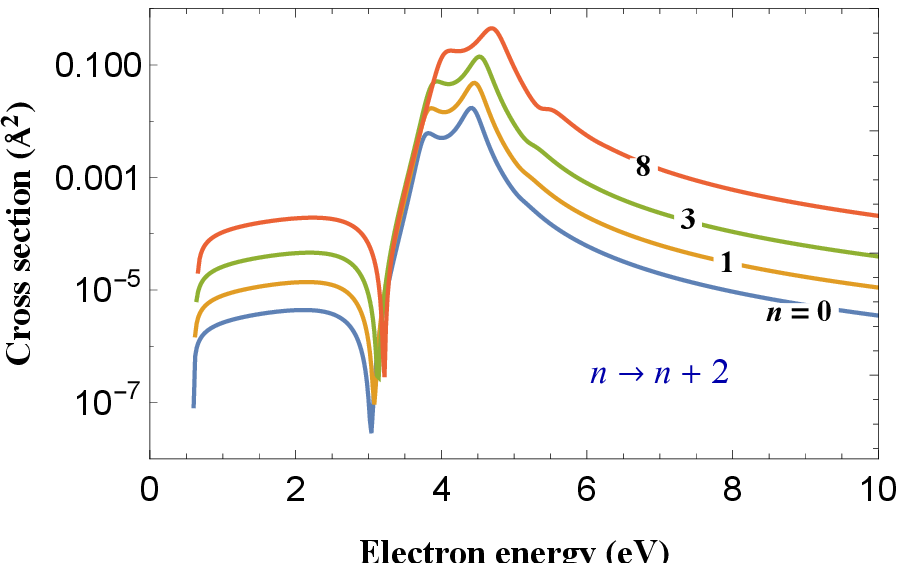}\\
\end{tabular}
\caption{Summary of the electron-CO$_2$ vibrational excitation cross sections as a function of the incident electron energy for the three normal modes. Upper panels: excitation cross section for processes starting from the lowest vibrational levels. Lower panels: mono-quantum (symmetric stretching) and double-quantum (bending and asymmetric stretching) cross sections for processes starting also from vibrationally excited molecules. \label{fig:CO2xsec}}
\end{figure}
\begin{figure}
\begin{tabular}{ccc}
Symmetric stretching & Bending mode & Asymmetric stretching\\
\includegraphics[scale=.6]{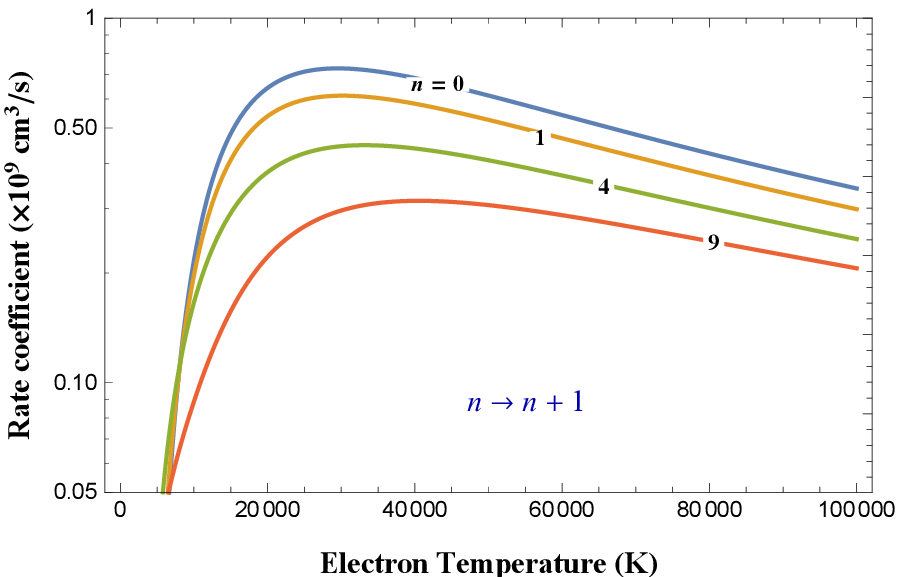}&\includegraphics[scale=.6]{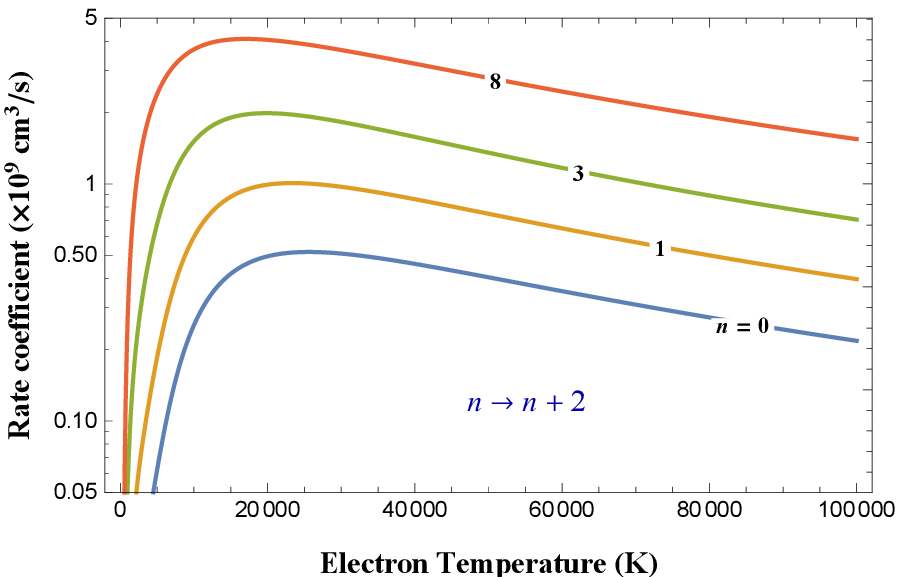}&\includegraphics[scale=.6]{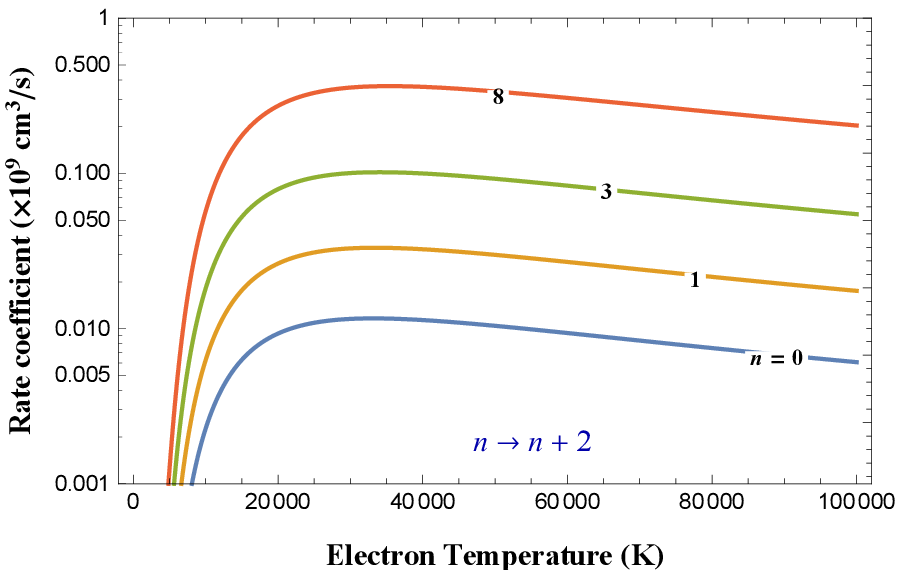}\\
\end{tabular}
\caption{Summary of selected electron-CO$_2$ vibrational excitation rate coefficients as a function of the electron temperature for the three normal modes. \label{fig:CO2rate}}
\end{figure}
Figure~\ref{fig:CO2xsec} shows selected cross section results, as a
function of the incident electron energy, for the vibrational
transitions described in process (\ref{process}), starting from the
ground vibrational levels and exciting the $n$-th level of
the three normal modes (upper panels). All cross section curves show a
main pronounced peak close to 3.8 eV which corresponds to the CO$^-_2$
resonance threshold. The other secondary peaks correspond to the
CO$_2^-$ vibrational levels (boomerang oscillations). The figures show
that the elastic cross sections ($n=0$) for the bending and asymmetric
stretching reach comparable values, an order of magnitude larger than
those for the symmetric stretch. The inelastic cross sections decrease
for increasing $n$, as expected, for all the three cases, also by
orders of magnitude. The same figure (lower panels) also shows the
cross sections for one-quantum transitions of the symmetric stretch motion, and
\textcolor{green}{two-quantum transitions occurring in the other two normal modes,
for which the selection rule, $\Delta n =0, 2$, holds. This is due to
the symmetric shape of the potentials and widths in bending and
antisymmetric modes, so that the Frank-Condon overlap between wave functions
having opposite parity, is suppressed.} \textcolor{blue}{Experimental investigations, actually, show that this is not the case. This aspect however is not fully clarified in literature. In fact, a coupled model could produce non-zero bending cross sections for odd-parity transitions as showed in Estrada \emph{et al} \cite{estrada:152} but also other processes, involving Feshbach resonances, could be important \cite{0953-4075-29-8-018}.} The cross sections shown in the lower panels of Fig.~\ref{fig:CO2xsec} refer to excitation process
starting also from vibrationally excited molecules. Figure~\ref{fig:CO2rate} shows some series of the rate
coefficients as a function of the electron temperature for the three
normal modes.

\begin{figure}
\includegraphics[scale=.7]{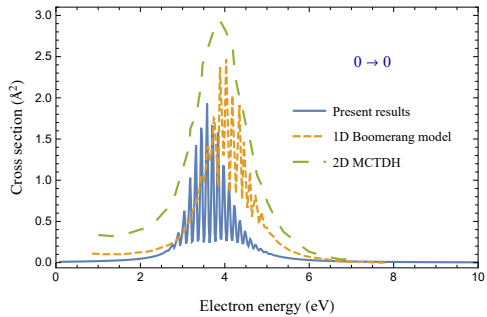}
\caption{Comparison between present result (solid blue curve) with the theoretical calculations of Rescigno \emph{et al.}~\cite{PhysRevA.65.032716} (broken curves) for elastic symmetric stretching $0\to0$ cross section. \label{fig:xsec_symm00_cfr}}
\end{figure}

\begin{figure}
\begin{tabular}{ccc}
\includegraphics[scale=.6]{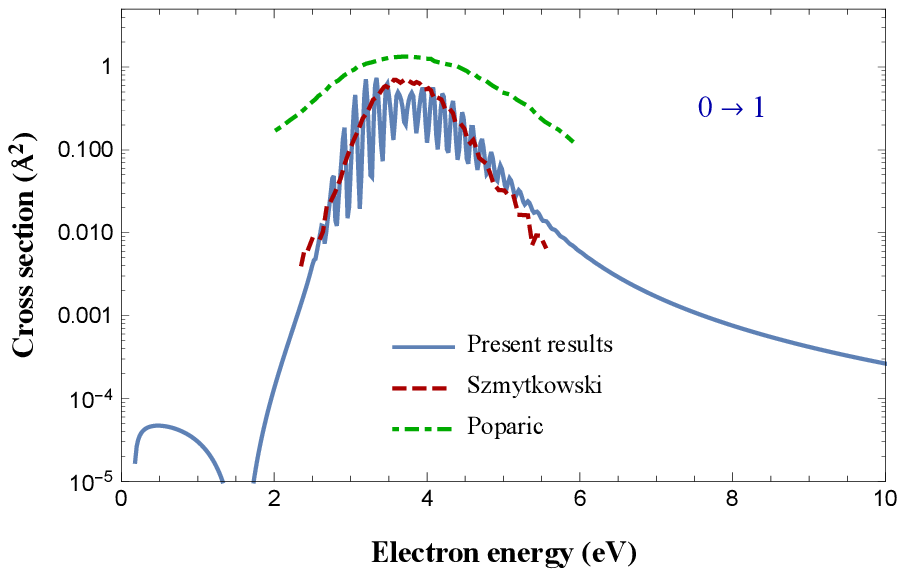} & \includegraphics[scale=.6]{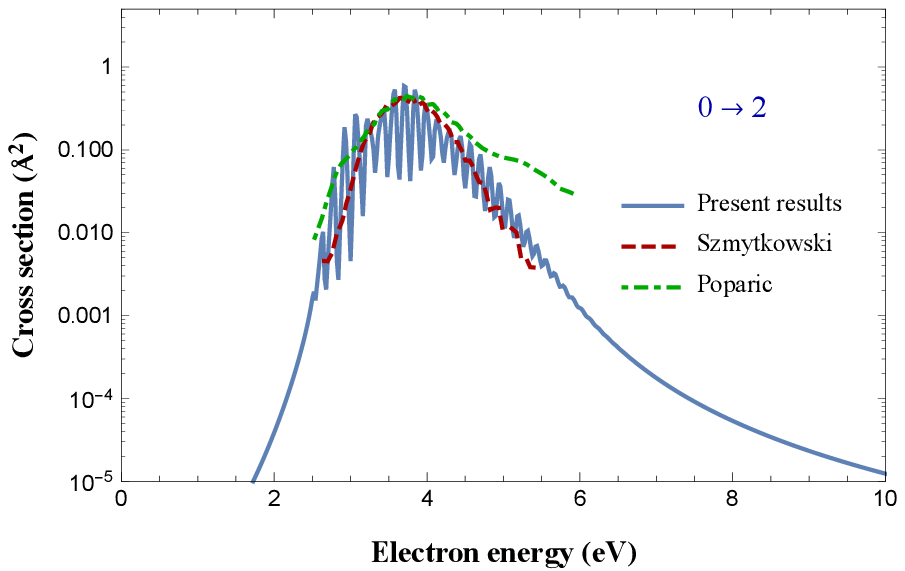} & \includegraphics[scale=.6]{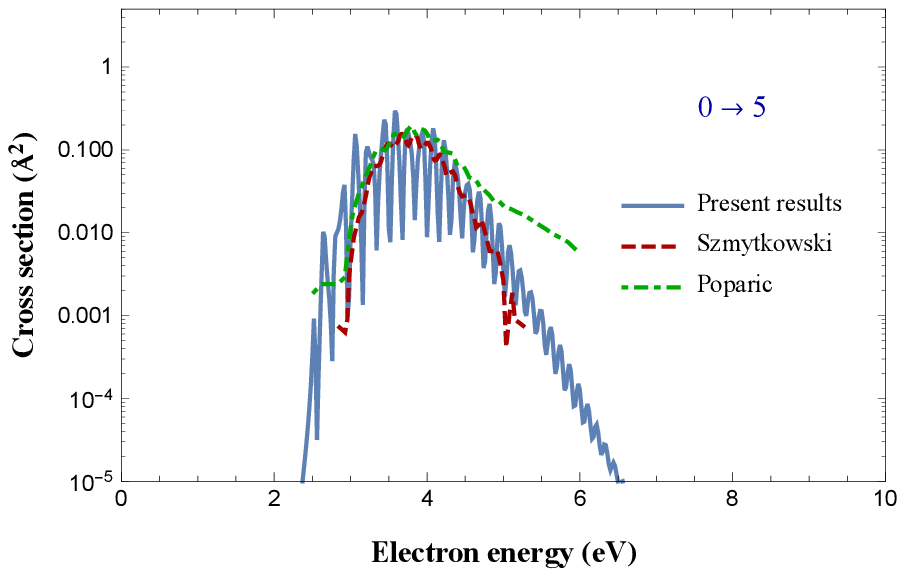}\\
\includegraphics[scale=.6]{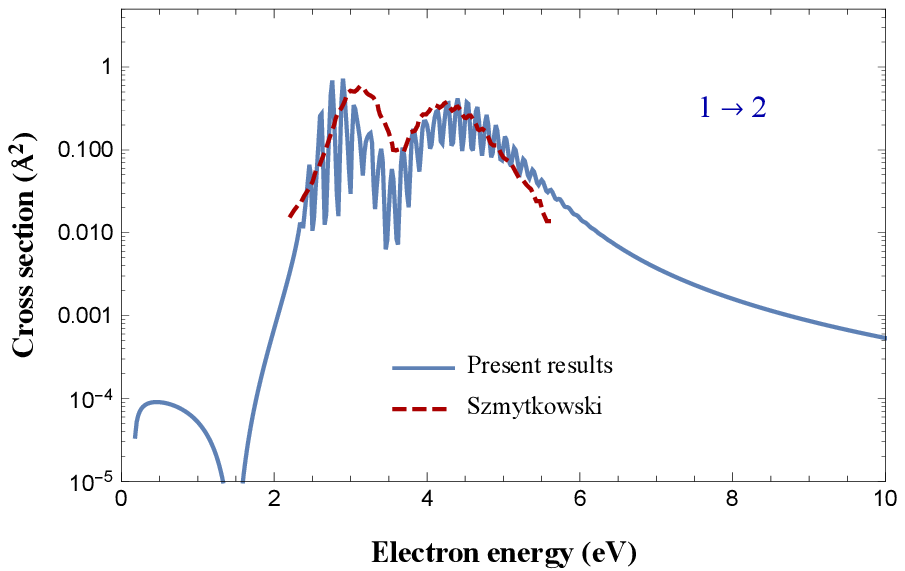} & \includegraphics[scale=.6]{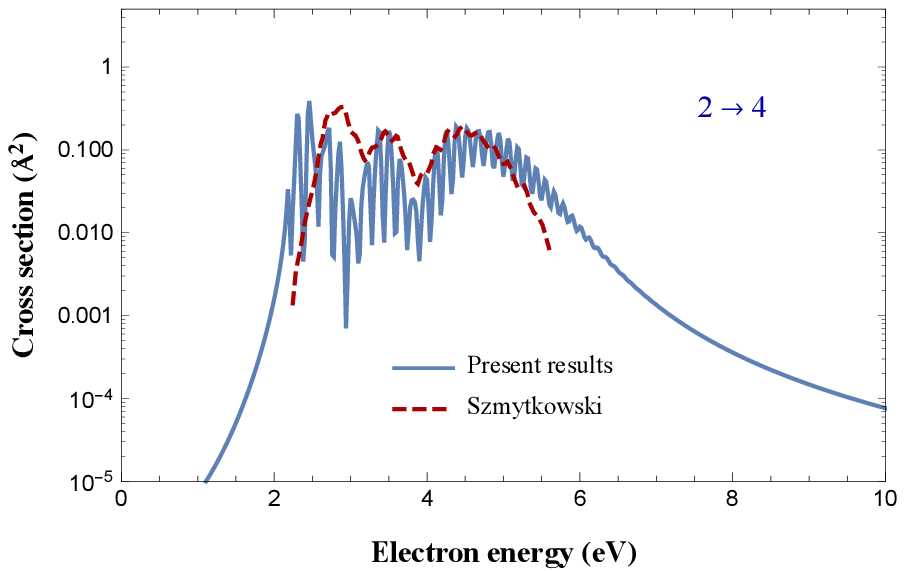} & \includegraphics[scale=.6]{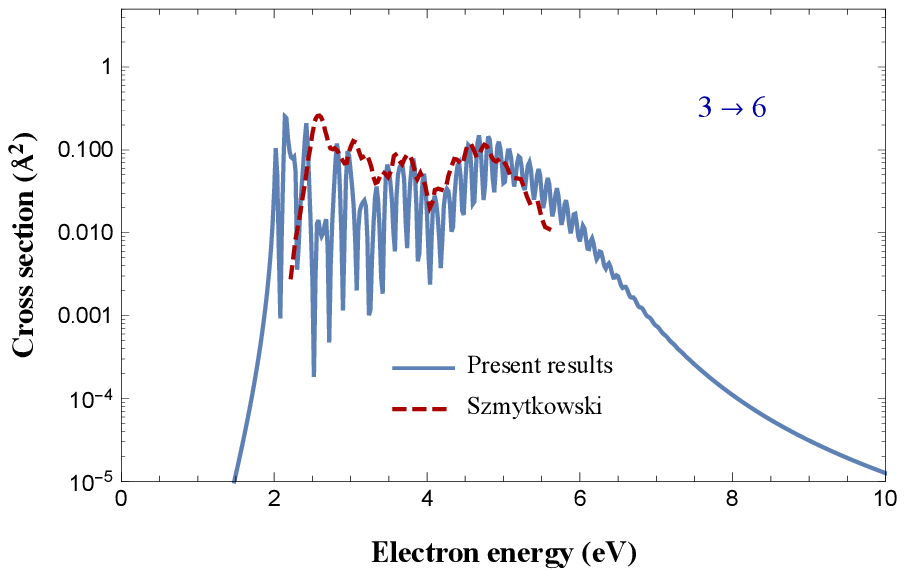}
\end{tabular}
\caption{Comparison between present results with available experimental data in Refs.~\cite{doi:10.1021/jp908593e, 0022-3700-11-12-004} for selected symmetric stretching cross sections. \label{fig:xsec_symm_cfr}}
\end{figure}

\begin{figure}
\begin{tabular}{cc}
\includegraphics[scale=.7]{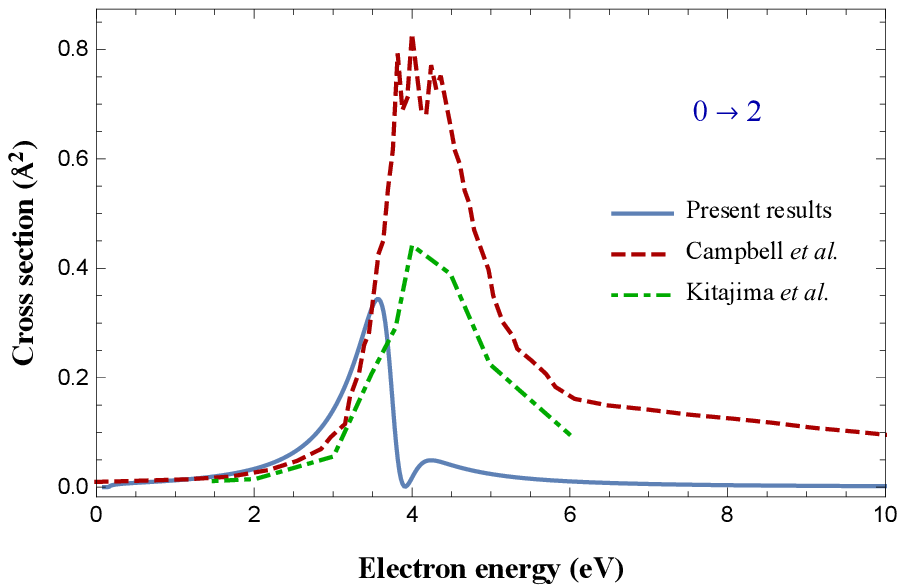} & \includegraphics[scale=.7]{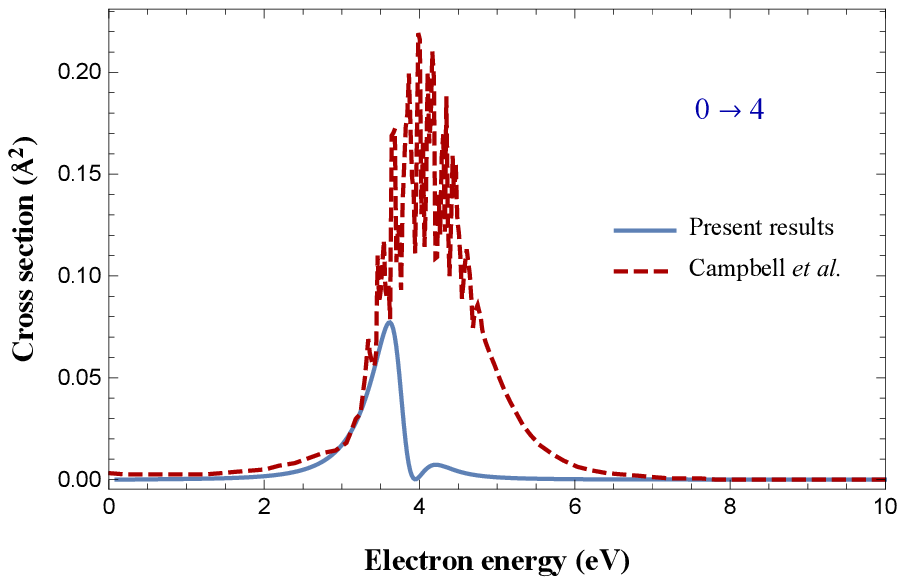}
\end{tabular}
\caption{Comparison between present results with available data in literature~\cite{JGRE:JGRE2501, KitajimaWatanabeTanakaEtAl2000} for bending stretching cross sections. \label{fig:xsec_bend_cfr}}
\end{figure}

Figure \ref{fig:xsec_symm00_cfr} shows the comparison with the theoretical calculations of Rescigno \emph{et al.}~\cite{PhysRevA.65.032716} for the elastic symmetric stretching $0\to0$ cross section. The present result as well as Rescigno's 1D calculation shows the well known resonant `boomerang' structure and they are in general agrement with each other. The slight shift of the peak positions is due to the different potential energy curves used in the calculations. On the other hand, Fig. \ref{fig:xsec_symm_cfr} shows very good agrement with experimental data available in literature for symmetric stretching cross sections. In the comparison for $0\to2$ bending mode transition cross sections, in Fig. \ref{fig:xsec_bend_cfr}, we find qualitative agreement with Kitajima \emph{et al.} results \cite{KitajimaWatanabeTanakaEtAl2000} and disagreement with the estimated data presented by Campbell \emph{et al.} \cite{JGRE:JGRE2501}. These last data, as Campbell \emph{et al.} state in their article, take into account also the Fermi dyad and triad coupling \cite{PhysRevA.67.042708}. However, the disagreement between the theoretical and experimental cross sections shows that further studies are needed to achieve reliable results  for the excitation of bending modes.

\textcolor{red}{In conclusion, in this Letter we presented a set of vibrational excitation cross sections for electron-CO$_2$ scattering in the separations of the mode approximation. In particular, we studied the role of the resonant contribution to the scattering. Good agreement, with respect to the data reported in literature, is found for the symmetric stretching mode whereas for the bending and asymmetric modes further investigations, taking into account non-resonant contributions, are needed.} Finally, the full set of data can be downloaded from the Phys4Entry database~\cite{F4Edatabase}.

\begin{acknowledgements}
The authors wish to thank Prof. M. Capitelli (University of Bari, Italy) for useful discussions and Prof. M. Panesi (University of Illinois, IL) for careful reading of the manuscript. One of the authors (VL) would like to thank the Politecnico di Bari for the kind hospitality during the preparation of the present article. This work received funding from the project ``Apulia Space'', PON 03PE-00067.6 from DTA Brindisi (Italy).
\end{acknowledgements}

\section*{References}
\addcontentsline{toc}{section}{References}

\bibliographystyle{is-unsrt}{}  %  <----

%\bibliography{C:/Users/vincenzo/Desktop/pandora/resonant}

\begin{thebibliography}{10}

\bibitem{Taylan_Berberoglu_2015}
O~Taylan and H~Berberoglu.
\newblock {\em Plasma Sources Sci. Technol.}, 24:\penalty0 015006, 2015.

\bibitem{Goede_et_al_2014}
Adelbert~P.H. Goede, Waldo~A. Bongers, Martijn~F. Graswinckel, Richard~M.C.M
  van~de Sanden, Martina Leins, Jochen Kopecki, Andreas Schulz, and Mathias
  Walker.
\newblock {\em EPJ Web of Conferences}, 79:\penalty0 01005, 2014.

\bibitem{doi:10.1021/acs.jpca.6b01154}
L.D. Pietanza, G.~Colonna, V.~Laporta, R.~Celiberto, G.~D’Ammando,
  A.~Laricchiuta, and M.~Capitelli.
\newblock {\em The Journal of Physical Chemistry A}, 120\penalty0
  (17):\penalty0 2614--2628, 2016.

\bibitem{0963-0252-24-1-015024}
TomÃƒÂ¡Ã…Â¡ KozÃƒÂ¡k and Annemie Bogaerts.
\newblock {\em Plasma Sources Science and Technology}, 24\penalty0
  (1):\penalty0 015024, 2015.

\bibitem{doi:10.1021/jp408522m}
Andrea Lombardi, Noelia Faginas-Lago, Leonardo Pacifici, and Alessandro
  Costantini.
\newblock {\em The Journal of Physical Chemistry A}, 117\penalty0
  (45):\penalty0 11430--11440, 2013.

\bibitem{0963-0252-23-4-045004}
TomÃ¡Å¡ KozÃ¡k and Annemie Bogaerts.
\newblock {\em Plasma Sources Science and Technology}, 23\penalty0
  (4):\penalty0 045004, 2014.

\bibitem{doi:10.1021/jp509843e}
André Janeco, Nuno~R. Pinhão, and Vasco Guerra.
\newblock {\em The Journal of Physical Chemistry C}, 119\penalty0 (1):\penalty0
  109--120, 2015.

\bibitem{PhysRevA.15.2186}
Michael~A. Morrison, Neal~F. Lane, and Lee~A. Collins.
\newblock {\em Phys. Rev. A}, 15:\penalty0 2186--2201, Jun 1977.

\bibitem{0022-3700-10-18-034}
I~Cadez, F~Gresteau, M~Tronc, and R~I Hall.
\newblock {\em Journal of Physics B: Atomic and Molecular Physics}, 10\penalty0
  (18):\penalty0 3821, 1977.

\bibitem{PhysRevLett.87.033201}
M.~Allan.
\newblock {\em Phys. Rev. Lett.}, 87:\penalty0 033201, Jun 2001.

\bibitem{PhysRevLett.80.1873}
L.~A. Morgan.
\newblock {\em Phys. Rev. Lett.}, 80:\penalty0 1873--1875, Mar 1998.

\bibitem{jt238}
J.~Tennyson and L.~A. Morgan.
\newblock {\em Philosophical Transactions of the Royal Society of London A:
  Mathematical, Physical and Engineering Sciences}, 357\penalty0
  (1755):\penalty0 1161--1173, 1999.

\bibitem{PhysRevA.64.040701}
Stephane Mazevet, Michael~A Morrison, Lesley~A. Morgan, and Robert~K. Nesbet.
\newblock {\em Phys. Rev. A}, 64:\penalty0 040701, Sep 2001.

\bibitem{itikawa:749}
Yukikazu Itikawa.
\newblock {\em Journal of Physical and Chemical Reference Data}, 31\penalty0
  (3):\penalty0 749--767, 2002.


\bibitem{estrada:152}
H.~Estrada, L.~S. Cederbaum, and W.~Domcke.
\newblock {\em The Journal of Chemical Physics}, 84\penalty0 (1):\penalty0
  152--169, 1986.



\bibitem{0963-0252-21-5-055018}
V~Laporta, R~Celiberto, and J~M Wadehra.
\newblock {\em Plasma Sources Science and Technology}, 21\penalty0
  (5):\penalty0 055018, 2012.

\bibitem{0963-0252-23-6-065002}
V~Laporta, D~A Little, R~Celiberto, and J~Tennyson.
\newblock {\em Plasma Sources Science and Technology}, 23\penalty0
  (6):\penalty0 065002, 2014.

\bibitem{PhysRevA.91.012701}
V.~Laporta, R.~Celiberto, and J.~Tennyson.
\newblock {\em Phys. Rev. A}, 91:\penalty0 012701, Jan 2015.

\bibitem{Celiberto_et_al_OPPJ_2014}
R.~Celiberto, V.~Laporta, A.~Laricchiuta, J.~Tennyson, and J.M. Wadehra.
\newblock {\em The Open Plasma Physics Journal, 2014, 7, (Suppl 1: M2) 33-47},
  7, (Suppl 1: M2):\penalty0 33--47, 2014.

\bibitem{PhysRevA.65.032716}
T.~N. Rescigno, W.~A. Isaacs, A.~E. Orel, H.-D. Meyer, and C.~W. McCurdy.
\newblock {\em Phys. Rev. A}, 65:\penalty0 032716, Feb 2002.

\bibitem{PhysRevA.67.042708}
C.~W. McCurdy, W.~A. Isaacs, H.-D. Meyer, and T.~N. Rescigno.
\newblock {\em Phys. Rev. A}, 67:\penalty0 042708, Apr 2003.

\bibitem{Herzberg_III}
Gerhard Herzberg.
\newblock {\em Molecular Spectra and Molecular Structure III. Electronic
  Spectra and Electronic Structure od Polyatomic Molecules}.
\newblock 1966.

\bibitem{MOLPRO_brief}
H.-J. Werner, P.~J. Knowles, G.~Knizia, F.~R. Manby, M.~{Sch\"{u}tz}, et~al.
\newblock {MOLPRO}, version 2010.1, a package of ab initio programs, 2010.

\bibitem{Tennyson_PR_2010}
Jonathan Tennyson.
\newblock {\em Physics Reports}, 491\penalty0 (2-3):\penalty0 29 -- 76, 2010.

\bibitem{jt518}
J.M. Carr, P.G. Galiatsatos, J.D. Gorfinkiel, A.G. Harvey, M.A. Lysaght,
  D.~Madden, Z.~Ma{\v s}{\'\i}n, M.~Plummer, J.~Tennyson, and H.N. Varambhia.
\newblock {\em The European Physical Journal D}, 66\penalty0 (3):\penalty0
  1--11, 2012.

\bibitem{Tennyson1984421}
Jonathan Tennyson and Cliff~J. Noble.
\newblock {\em Computer Physics Communications}, 33\penalty0 (4):\penalty0 421
  -- 424, 1984.

\bibitem{0034-4885-31-2-302}
J~N Bardsley and F~Mandl.
\newblock {\em Reports on Progress in Physics}, 31\penalty0 (2):\penalty0 471,
  1968.


\bibitem{0953-4075-29-8-018}
David~C Cartwright and Sandor Trajmar.
\newblock {\em Journal of Physics B: Atomic, Molecular and Optical Physics},
  29\penalty0 (8):\penalty0 1549, 1996.


\bibitem{doi:10.1021/jp908593e}
G.~B. Popari{\'c}, M.~M. Risti{\'c}, and D.~S. Beli{\'c}.
\newblock {\em The Journal of Physical Chemistry A}, 114\penalty0 (4):\penalty0
  1610--1615, 2010.

\bibitem{0022-3700-11-12-004}
C~Szmytkowski, M~Zubek, and J~Drewko.
\newblock {\em Journal of Physics B: Atomic and Molecular Physics}, 11\penalty0
  (12):\penalty0 L371, 1978.

\bibitem{JGRE:JGRE2501}
L.~Campbell, M.~J. Brunger, and T.~N. Rescigno.
\newblock {\em Journal of Geophysical Research: Planets}, 113\penalty0
  (E8):\penalty0 n/a--n/a, 2008.


\bibitem{KitajimaWatanabeTanakaEtAl2000}
M.~Kitajima, S.~Watanabe, H.~Tanaka, M.~Takekawa, M.~Kimura, and Y.~Itikawa.
\newblock {\em Phys. Rev. A}, 61:\penalty0 060701, May 2000.

\bibitem{F4Edatabase}
Database of the european union phys4entry project, 2012-2016.

\end{thebibliography}

\end{document}